\documentclass{article}

\usepackage{arxiv}
\usepackage[T1]{fontenc}    
\usepackage{hyperref}       
\usepackage{url}            
\usepackage{booktabs}       
\usepackage{amsfonts}       
\usepackage{nicefrac}       
\usepackage{microtype}      
\usepackage{graphicx}
\usepackage[square,sort,comma,numbers]{natbib}
\usepackage{doi}
\usepackage{caption}
\usepackage{subcaption}
\usepackage{booktabs, tabularx}
\usepackage{xcolor}
\usepackage{multirow}
\usepackage{anyfontsize}
\usepackage{verbatim}

\usepackage{xcolor}
\PassOptionsToPackage{usenames,dvipsnames}{xcolor}

\title{Information Extraction from Historical Well Records Using A Large Language Model}

\author{{\hspace{1mm}Zhiwei Ma\textsuperscript{1,}\thanks{Corresponding author, email: m.ma@lanl.gov}}
	\And
	{\hspace{1mm}Javier E. Santos\textsuperscript{1}}
	\And
        {\hspace{1mm}Greg Lackey\textsuperscript{2}} 
 	\And
	{\hspace{1mm}Hari Viswanathan\textsuperscript{1}} 
	\And
        {\hspace{1mm}Daniel O'Malley\textsuperscript{1}}
        \And
        {\textsuperscript{1} \normalfont The Earth and Environmental Sciences Division, 
        Los Alamos National Laboratory,  
        Los Alamos, NM 87544}
        \And
        {\textsuperscript{2} \normalfont Geological \& Environmental Systems,
        National Energy Technology Laboratory, 
        Pittsburgh, PA 15236}
}

\renewcommand{\shorttitle}{Information Extraction from Historical Well Records using A Large Language Model}

\hypersetup{
pdftitle={A template for the arxiv style},
pdfsubject={q-bio.NC, q-bio.QM},
pdfauthor={David S.~Hippocampus, Elias D.~Striatum},
pdfkeywords={First keyword, Second keyword, More},
}

\begin{document}
\maketitle

\begin{abstract}
To reduce environmental risks and impacts from orphaned wells (abandoned oil and gas wells), it is essential to first locate and then plug these wells. Although some historical documents are available, they are often unstructured, not cleaned, and outdated. Additionally, they vary widely by state and type. Manual reading and digitizing this information from historical documents are not feasible, given the high number of wells. Here, we propose a new computational approach for rapidly and cost-effectively locating these wells. Specifically, we leverage the advanced capabilities of large language models (LLMs) to extract vital information including well location and depth from historical records of orphaned wells. In this paper, we present an information extraction workflow based on open-source Llama 2 models and test them on a dataset of 160 well documents. Our results show that the developed workflow achieves excellent accuracy in extracting location and depth from clean, PDF-based reports, with a 100\% accuracy rate. However, it struggles with unstructured image-based well records, where accuracy drops to 70\%. The workflow provides significant benefits over manual human digitization, including reduced labor and increased automation. In general, more detailed prompting leads to improved information extraction, and those LLMs with more parameters typically perform better. We provided a detailed discussion of the current challenges and the corresponding opportunities/approaches to address them. Additionally, a vast amount of geoscientific information is locked up in old documents, and this work demonstrates that recent breakthroughs in LLMs enable us to unlock this information more broadly.
\end{abstract}



\section{Introduction}
\label{sec:intro}
In the oil and gas industry, orphaned wells are defined as a class of unplugged wells whose owner/operator is unknown. Thus, other than agencies from the government, no one is responsible for the well-plugging operations and site restoration processes \cite{boutot2022documented,dan2024OW}. While some orphaned wells are well-documented with detailed information, such as name, location, and drilling details, many others lack important information and are referred to as undocumented orphaned wells. Based on a recent report from the U.S. Geological Survey (USGS), there are only 117,672 documented orphaned oil and gas wells in the 27 states in the U.S. \cite{merrill2023analysis}. On the other hand, the Interstate Oil and Gas Compact Commission (IOGCC) reported there are between 310,000 to 800,000 undocumented orphan wells in the 32 states of the U.S. that produce the most oil and gas, as of 2020 \citep{IOGCC2021}. However, it is believed that the actual number of undocumented orphan wells is much larger. Orphaned wells often present numerous environmental and health risks, including emitting methane, releasing  hazardous air pollutants, creating a risk of explosion, leaking continent to underground water \cite{methane_2021,kang_environmental_2023,raimi_decommissioning_2021}. For example, according to \cite{kang_environmental_2023, methane_2021}, in the U.S., the methane emissions from all abandoned oil and gas wells amounted to about 3\% of those from natural gas and petroleum systems. However, the `documented' orphaned wells that are covered by the Bipartisan Infrastructure Law (BIL) only emit approximately 3\% to 6\% of total U.S. methane emissions from all abandoned oil and gas wells. Therefore, it is necessary to find vital information on the orphaned wells such as well locations and depths for subsequent treatments to mitigate these environmental risks. A detailed review of the challenges associated with orphaned wells can be found from O'Malley et al. (2024) \citep{dan2024OW}.

Oil and gas regulatory agencies in the U.S. maintain regulatory records (e.g., permitting documents) for wells under their jurisdiction that often contain valuable information about the location and construction of wells. These historical records are often decades old and exist in a variety of formats that sometimes include digital PDFs but are usually scanned images or paper copies. The current practice for extracting information from historical documents related to orphaned wells involves hiring individuals to review and enter the data into a computer. This manual process requires some domain knowledge to accurately interpret the documents and correct errors, which are frequently compounded by the presence of stamps and various information formats (e.g., 45\textdegree25'28.56'' and 56.358599 degrees, when dealing with units of latitude). Given the high number of orphaned wells, it is neither practical nor realistic to manually extract and digitize this information from historical well documents. That is because the manual extraction process is labor-intensive and time-consuming. Therefore, it is crucial to develop an automatic information extraction workflow to analyze those historical well documents, facilitating the rapid and precise identification of the wells' location and depth information. To deal with this challenge, we developed an information extraction workflow combining text extraction techniques, (e.g., Optical Character Recognition or OCR) and large language models (LLMs). Specifically, OCR technology is used to convert different types of well documents such as PDFs and scanned images, into machine-encoded texts, which are editable and searchable data \cite{eikvil1993optical,chaudhuri2017optical}. Next, we employed publicly available pre-trained LLMs to perform the well information extraction, during which, the converted texts are used as inputs for a properly-designed prompt. This developed workflow is based on the strong capabilities of LLMs.

LLMs belong to artificial intelligence or machine learning and are pre-trained language models on vast amounts of data \cite{chang2023survey, topsakal_creating_2023}. Recently, artificial intelligence and machine learning have rapidly advanced and been widely adopted in the geoscience and subsurface flow fields for various applications. These include well control/production optimization in oil/gas applications \cite{ma2022optimization, ma2020design}, reconstruction of complex spatial fields for geospatial analysis \cite{santos2023development}, for upscaling geomechanical properties \cite{zhang2023upscaling}, for geological CO\textsubscript{2} storage modeling \cite{yan2022improving, ma2023CO2}, for rapid forecasting and history matching in unconventional reservoirs \cite{srinivasan2021machine}, and for inference of random medium properties \cite{10440111}. As one type of artificial intelligence models, LLMs can be described as extensive, pre-trained statistical language models that utilize neural networks \cite{minaee2024large}. The development and advances in LLMs are very fast. These developments include the introduction of new models and increased model parameter sizes, along with incorporating domain information for fine-tuned LLMs. New fine-tuned versions of base models are released many times per day, and new base models such as Llama, Mistral, and Mixtral are also released frequently. Currently, many LLMs are referred to transformer-based neural language models. These models typically possess billions of parameters and are trained using an extremely large dataset \cite{minaee2024large}. Due to their emergent ability and generalizability \cite{pan2024unifying}, LLMs are capable of generating text, understanding natural language, translating, summarizing content, and performing sentiment analysis, among other capabilities. Examples of applications of LLMs can be found in the following categories: translation \cite{koshkin2024transllama}, sentiment analysis \cite{sun2023sentiment}, question and answering \cite{NEURIPS2023_9cb2a749}, code generation \cite{wang2021codet5}, summarization \cite{shekhar2024optimizing}, and chatbots \cite{tan2024finetuning}. In the field of hydrology and earth science, a brief overview of opportunities, prospects, and concerns using ChatGPT was provided \cite{foroumandi2023chatgpt}. 
The research topic addressed in this work pertains to question-answering category. In other words, we pose specific questions to the LLMs based on well records and anticipate that the LLMs will generate the desired answers, after analyzing the provided text. Our objective is to leverage LLMs's capability for processing text as an alternative approach to overcome challenges associated with the manual extraction of well information from historical documents, as highlighted above.

In this work, we mainly focused on the Llama 2 family of Large Language Models. Llama 2 is an updated version of Llama 1 \cite{touvron2023llama} and it was trained on a mix of data that are publicly available \cite{touvron2023llama2}. In addition, there is a 40\% increase in the pre-training corpus, with the model's the context length being doubled when compared with Llama 1. Meta’s release of Llama 2 family consists of several pre-trained Llama 2 models, ranging from 7 billion to 70 billion parameters, along with their corresponding fine-tuned LLMs for dialogue use cases. Training Llama 2 models is not trivial, as they require advanced GPU clusters. To train these models, Meta used two clusters equipped with NVIDIA A100 GPUs. It took about 3,311,616 GPU hours to train these models and with 539 tCO\textsubscript{2}eq generated. According to \cite{touvron2023llama2}, Llama 2-chat models, in general, have a better performance than some open-source models on a series of safety and helpfulness benchmark tests. In addition to that, the authors also claimed that Llama 2 models achieve performance comparable to some closed-source models in their human evaluations. Because of their superior performance and open-source nature, Llama 2 models were used for analysis in this work.

In order to interact with LLMs and receive responses, it is common to use prompts \cite{chang2023survey}. A typical prompt consists of three elements: instruction, context, and input text \cite{pan2024unifying}. As a new field, the goal of prompt engineering to improve LLMs performance for a given task by creating and refining prompt contents \cite{pan2024unifying}. Recently, various prompting approaches have been developed to improve the reasoning capability of LLMs \cite{huang2022large}. One of these examples is the chain-of-thought strategy proposed by \cite{wei2022chain}, in which the LLMs are asked to provide a series of intermediate reasoning steps and to improve the final performances for complex reasoning tasks \cite{wei2022chain, huang2022large}. In this work, we optimized prompt contents including the approach of chain-of-thought in order to improve the performance of well information extraction tasks.

The contribution of this work can be summarized as follows: First, we developed a new LLM-based workflow for well information extraction. To the best of our knowledge, this paper is the first of its kind to practically employ LLMs for extracting vital information pertinent to managing orphaned oil and gas wells. Therefore, this work could serve as a significant example for information identification tasks for other researchers and the research communities. Second, we conducted a detailed analysis of the impact of prompts, model sizes, and the chain-of-thought strategy on the extraction performance, aspects previously unexplored in the field of geoscience. Third, the developed workflow can be easily deployed and we believe that employing this workflow can significantly accelerate information digitization from historical well documents.

The rest of this paper is organized as follows: we will first introduce our detailed methodology related to the workflow of information extraction, historical well records, text extraction, the theory of LLMs, Llama 2, and performance evaluations in Section \ref{sec:method}. We will present the extraction results including various treatments that are incorporated in this work in Section~\ref{sec:results}, which is followed by a brief discussion of the challenges and the corresponding opportunities in Section \ref{sec:discussions}. Finally, we will summarize the major findings and provide the potential future works in Section \ref{sec:concluding-remarks}.
\section{Method}
\label{sec:method}
In this section, we provide a detailed description of the proposed information extraction workflow, historical well records, large language models used for information extractions, and performance evaluation methods.

\subsection{Information extraction workflow via large language models (LLMs)}
\begin{figure}
	\centering
        \includegraphics[width=10cm]{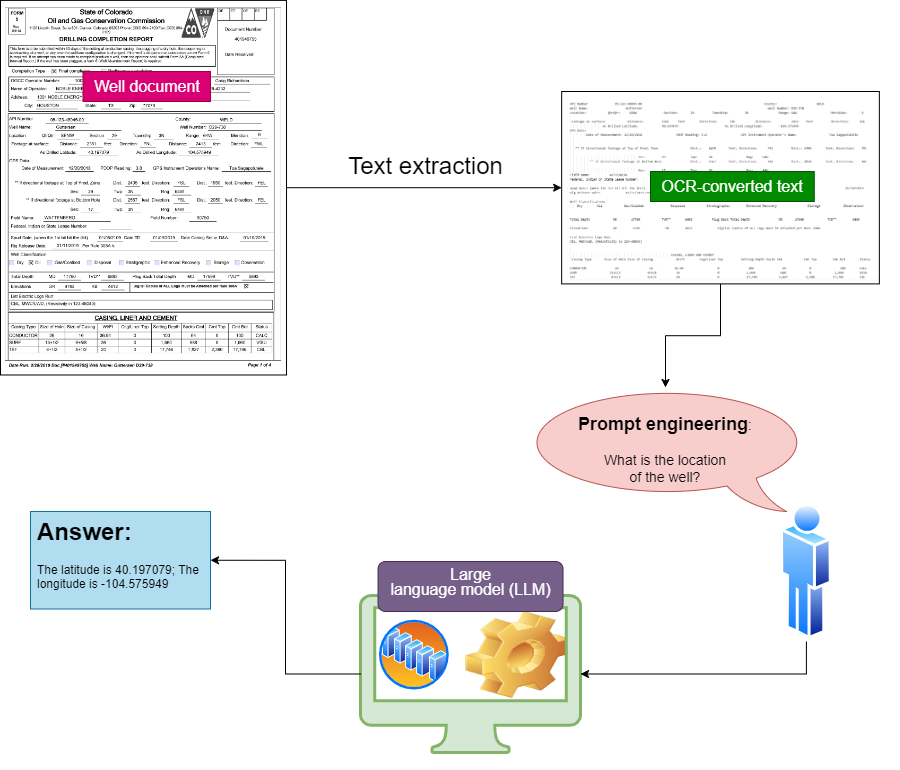}
	\caption{The proposed workflow for well information extraction via LLM.}
	\label{fig:workflow}
\end{figure}

The proposed workflow for information extraction from orphaned well historical records is presented in Figure~\ref{fig:workflow}. Given that LLMs can only process text information, the first step involves converting historical documents into text via text extraction approaches such as OCR. Next, the converted texts are subjected to LLMs. Here, we integrate the texts into some predefined prompt templates to form the final question prompt.

\begin{figure}
	\centering
        \includegraphics[width=8cm]{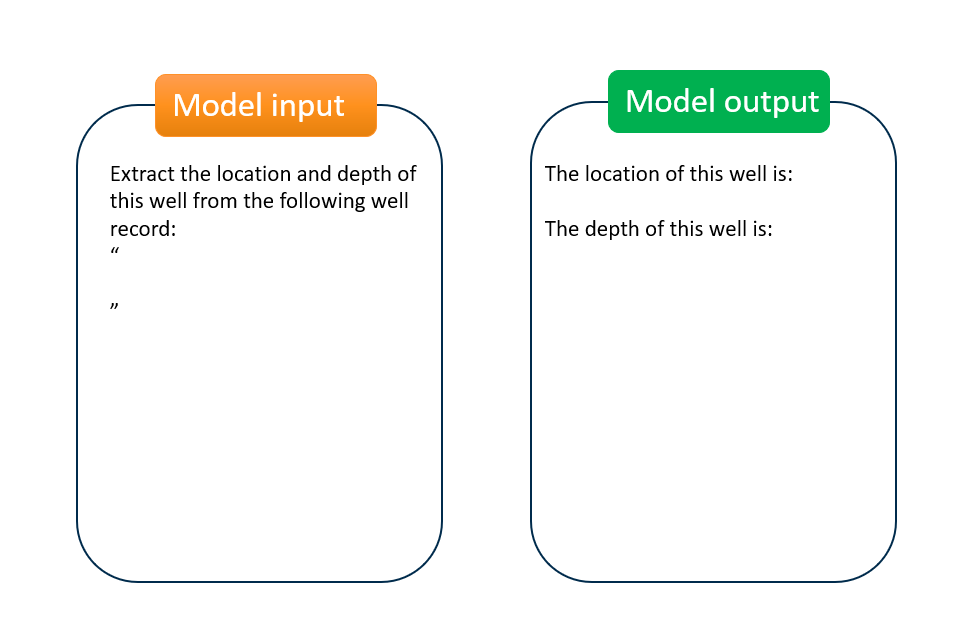}
	\caption{An illustration of model inputs and outputs for LLM. Note that we aim to show the structure of the model's input and output. One has to provide specific well record texts to the model input section, and the LLM would generate the corresponding detailed output in terms of well location and depth.}
	\label{fig:prompt_template}
\end{figure}

After running the LLMs with the complete prompt, the next step is generating an answer in text format, as shown in Figure \ref{fig:prompt_template}. The answers can be examined, and if the result is satisfactory, the information extraction task for this historical document is completed. Otherwise, we may need to refine the prompt or switch to different LLMs to achieve the desired outputs. In this work, we created a for-loop to automatically extract the information of interest from the well documents. The following subsections will cover the detailed methodologies for each major step in the information extraction workflow.

\subsection{Historical well records}
\label{subsec:Well-records}
\begin{figure}
     \centering
     \begin{subfigure}[b]{0.485\textwidth}
         \centering
         \includegraphics[width=\textwidth]{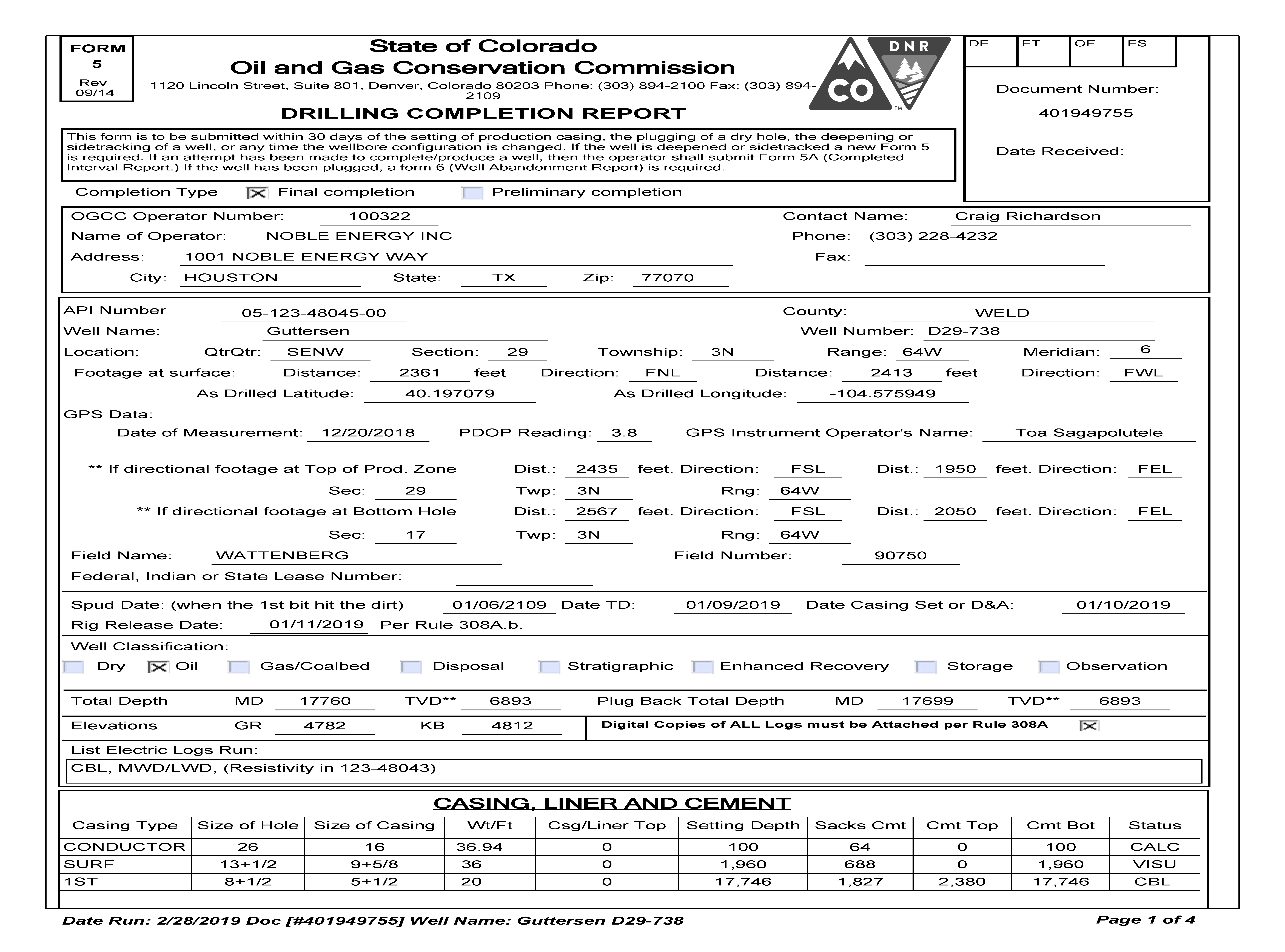}
         \caption{Colorado drilling completion report}
         \label{fig:well_record_examples_for_this_paper_a}
     \end{subfigure}
     \hfill
     \begin{subfigure}[b]{0.485\textwidth}
         \centering
         \includegraphics[width=\textwidth]{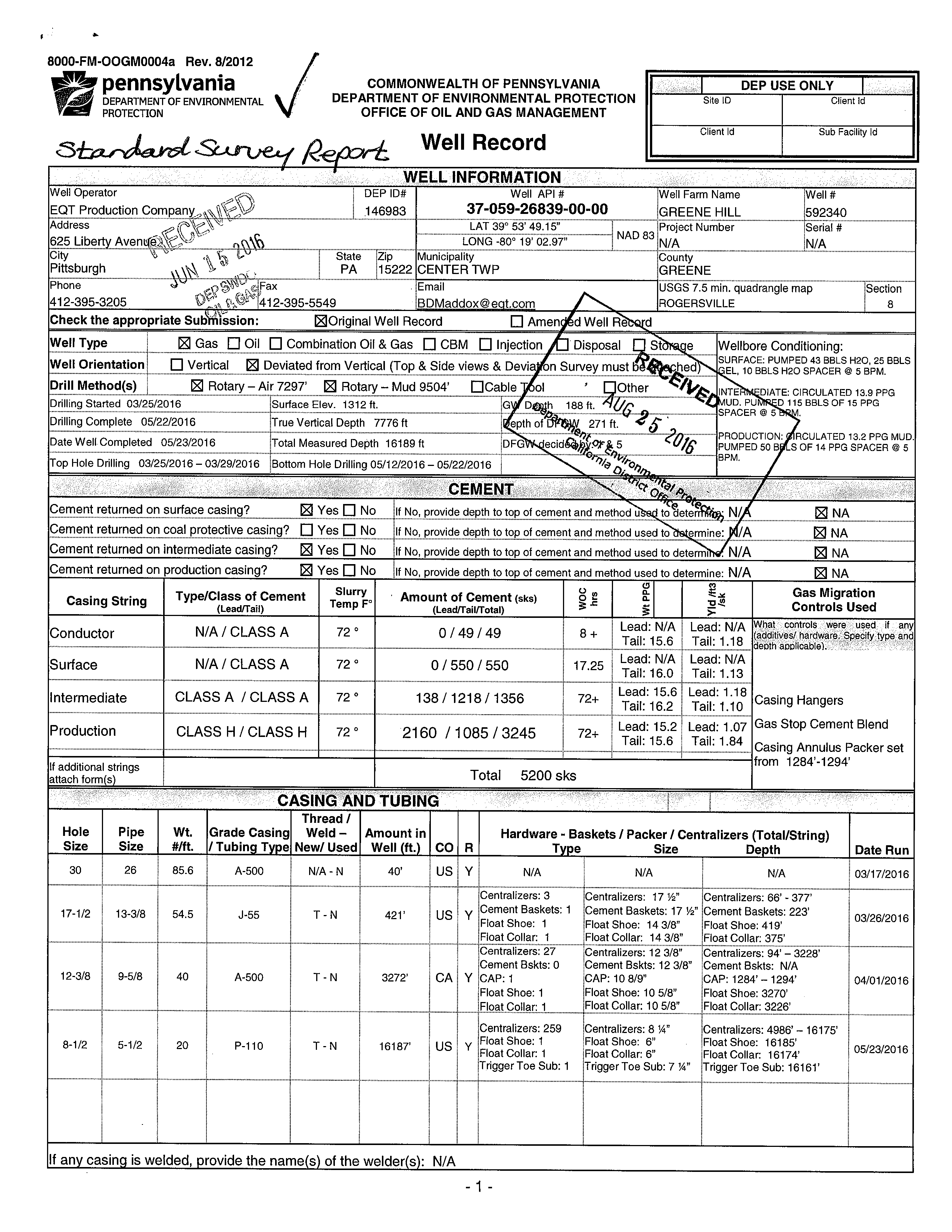}
         \caption{Pennsylvania well record}
        \label{fig:well_record_examples_for_this_paper_b}
     \end{subfigure}
        \caption{Examples of well records used in this study.}
        \label{fig:well_record_examples_for_this_paper}
\end{figure}

In this study, we analyze two types of well records: well drilling completion reports from Colorado and well record reports (Unconventional Operators) from Pennsylvania, as illustrated in Figure~\ref{fig:well_record_examples_for_this_paper}. These types of well records are commonly utilized by oil and gas regulatory agencies to document the construction history of wells. However, it is worth noting that each jurisdiction tracks well information with their own records that have unique formats. The multitude of oil and gas jurisdictions in the U.S. and the differences between the records they use increases the practical challenge of digitizing well information into a unified platform for characterizing orphaned wells. 

The preliminary dataset assembled for this study includes 150 well drilling completion reports from Colorado and 10 well records from Pennsylvania for demonstration purposes. The well records presented in Figure~\ref{fig:well_record_examples_for_this_paper} contain a wealth of information, such as the operator's name, address, and phone number; the American Petroleum Institute (API) number (a unique identifier assigned to each oil and gas well), name and location of the well; the spud date; the depth; and details on casing, liner, and cement. Although well records contain an abundance of information, the location and depth data are crucial for well remediation and will be extracted using LLMs in this work. That is because depth information provides a better understanding of the casing depth. 

As shown in Figure~\ref{fig:well_record_examples_for_this_paper}, we can see that well drilling completion reports from Colorado are clean and digitized. Well records from Pennsylvania contain many hand-written words and stamps. For example, in the top left corner, there are three hand-written words: "Standard Survey Report". There is a stamp on the mid-right side of this record, which shows "RECEIVED AUG 25 2016 Department of Environmental Protection California District Office". In addition, the middle part of the document is somewhat blurred with grey shadow. All these hand-written words, marks, and stamps increase the challenge of information extraction using LLMs. That is because the LLMs employed in this work require texts as input; therefore we must utilize text extraction technologies (e.g., OCR) to convert the image-based well records into text. 

\subsection{Text extraction} 
Plain text was acquired from the Colorado and Pennsylvania well records using two different approaches selected based on the original format of the document: 1) PDF to text conversion and 2) optical character recognition (OCR). Colorado well records were stored in text-based PDF format, which enabled a direct extraction of embedded text using the open-source tool \textit{pdftotext} \cite{pdftotext}. Pennsylvania well records were stored as scanned image files, which have no embedded text. Consequently, Google's Enterprise OCR, made available through their Document AI API \cite{googledocumentai}, was used to convert text in the Pennsylvania records into a machine-readable format. 

\begin{figure}
     \centering
     \begin{subfigure}[b]{0.65\textwidth}
         \centering
         \includegraphics[width=\textwidth]{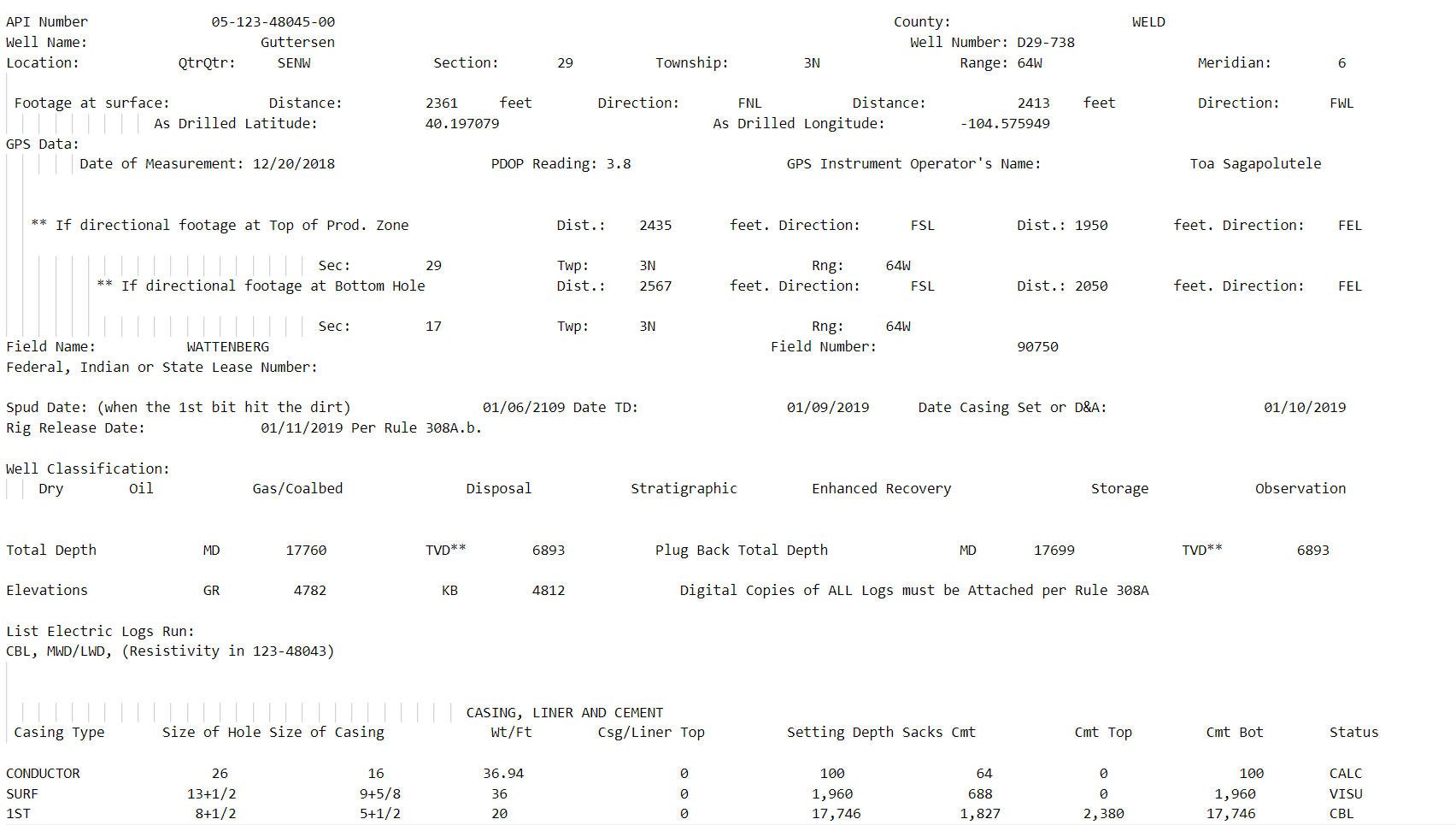}
         \caption{Colorado drilling completion report}
         \label{fig:well_record_text_examples_for_this_paper_a}
     \end{subfigure}
     \hfill
     \begin{subfigure}[b]{0.3\textwidth}
         \centering
         \includegraphics[width=\textwidth]{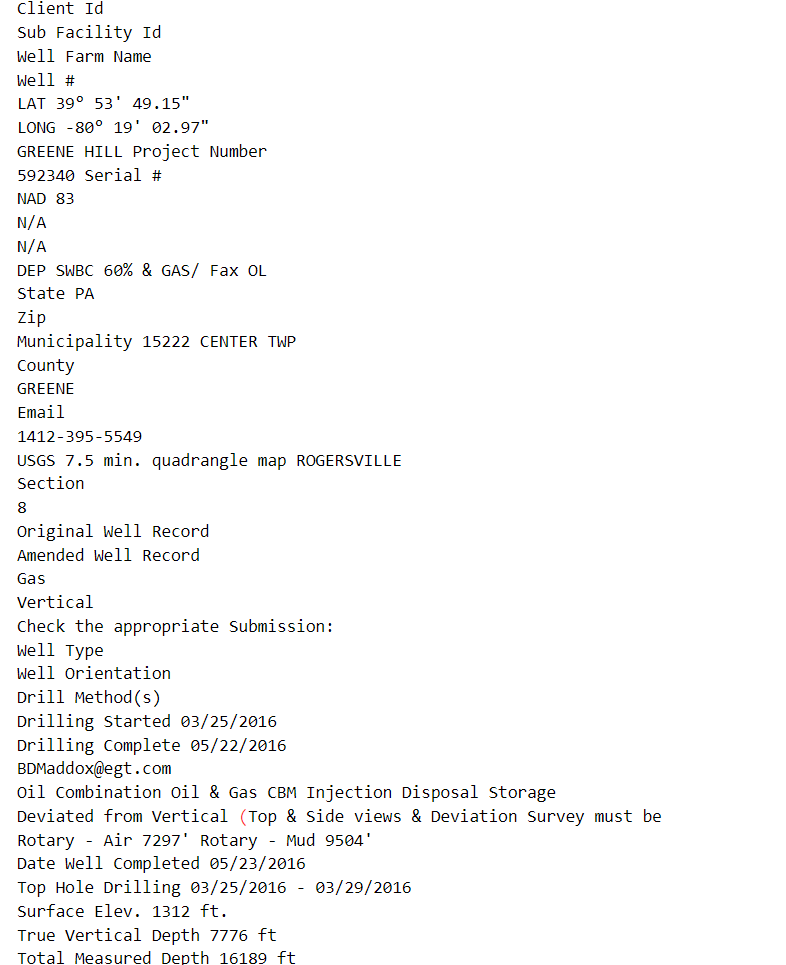}
         \caption{Pennsylvania well record}
         \label{fig:well_record_text_examples_for_this_paper_b}
     \end{subfigure}
        \caption{Part of the texts extracted from the two well records shown in Figure \ref{fig:well_record_examples_for_this_paper}.}
        \label{fig:well_record_text_examples_for_this_paper}
\end{figure}

Figure ~\ref{fig:well_record_text_examples_for_this_paper} displays a portion of the text information extracted from the two examples shown in Figure~\ref{fig:well_record_examples_for_this_paper}. When compared with the original documents in Figure~\ref{fig:well_record_examples_for_this_paper}, the quality of plain text extraction is acceptable, as the information presented in Figure~\ref{fig:well_record_examples_for_this_paper} match that in the original documents. For example, the converted information of well locations (latitude and longitude) agrees with that in the two documents in Figure~\ref{fig:well_record_examples_for_this_paper}. 
Another observation is that the formatting of converted information is not the same. The structure of the PDF converted text in Figure~\ref{fig:well_record_text_examples_for_this_paper_a} preserves the alignment and structure of the original document as in Figure~\ref{fig:well_record_examples_for_this_paper_a}. However, the OCR extracted text in Figure~\ref{fig:well_record_text_examples_for_this_paper_b} does not maintain a similar table-style structure to that in the well record shown
in Figure~\ref{fig:well_record_examples_for_this_paper_b}. Instead, the words within one single line in the image are divided into multiple rows in the OCR-processed text. The lack of correct structure in OCR-converted text poses a significant challenge for information extraction using LLMs as it requires LLMs to have advanced understanding capability to analyze the overall text. More advanced computer vision approaches should be developed and applied for text extraction to improve performance, which is beyond the scope of this paper. Once we have extracted the text information, the next step is to feed it into a pre-designed prompt for LLM for extracting well location and depth.

\subsection{Large language models (LLMs)} 
LLMs are machine learning models trained on vast amounts of text data. This training enables them to comprehend and produce text that closely resembles human writing. The sheer scale of these models, coupled with the large amounts of data they are trained on (on the order of trillions of tokens), allows them to learn complex patterns and relationships within the text. As the training progresses, these models develop abilities to perform a variety of tasks. For example, they can accurately answer queries, summarize vast amounts of information, and generate new text that is both coherent and contextually sound. 

Llama 2 \cite{touvron2023llama2}, developed by Meta AI, is a large language model that has attracted attention from the research community for its capabilities. It follows a structure similar to GPT (Generative Pretrained Transformer) \cite{radford2018improving}, which relies on stacked attention layers to process and generate text. These layers work by focusing on different parts of the input text to determine what is important and what is not. This mechanism enables Llama 2 to process and generate text effectively, understanding the context and nuances of the input text.

Available in different sizes, from smaller versions like Llama 2 7B to the largest, Llama 2 70B, these variants differ in their processing power and the depth of understanding they can provide. Larger models, while requiring more computational resources, can deliver more accurate and nuanced interpretations of data. Llama 2 operates under an open-weights regime, meaning the model weights are accessible to the public, but the specific data used for training these models is not disclosed. Llama 2 comes in two main versions: Foundational and Chat. The Foundational model is a general-purpose tool for text completion, while the Chat model has been further refined with techniques like supervised fine-tuning and Reinforcement Learning from Human Feedback (RLHF) to enhance its abilities to be a useful assistant. In our work, we focused on the assistant-type models, which are optimized for tasks requiring in-depth analysis and information extraction. 

As stated in the Introduction section, prompt engineering is a crucial aspect of working with chat models. It involves crafting the input text (or \textit{prompt}) to guide the model in generating a desired output. Through effective prompt engineering, users can steer the LLMs' focus and improve the quality of the extracted information. Moreover, LLMs like Llama 2 can perform zero-shot learning, which allows the model to make predictions or generate responses in tasks it has not explicitly been trained on. Contrarily, few-shot learning for LLMs refers to the process of a model learning from a small number of examples. For a specific task, we can provide providing a few demonstrative examples in the prompt to enhance the performance through few-shot learning. Another useful concept for LLM prompting is the \textit{chain-of-thought} approach \cite{wei2022chain}. This involves the model breaking down a problem into smaller, manageable parts, similar to how humans approach complex problems. This method can enhance the model's ability to understand and solve intricate tasks, making it a very useful approach for analyzing and extracting data from extensive and complex records. In this work, we mainly tested both zero-short learning and chain-of-thought methods. 

It is worth noting that we used quantized Llama 2 models in this work to reduce the memory footprint, which was needed by the hardware we used. Specifically, for example, for Llama 70B model, we utilized the Llama-2-70B-chat-GPTQ \cite{TheBloke-Llama-2-70B-chat-GPTQ} downloaded from the Hugging face \cite{wolf2019huggingface} due to the reduced size, instead of its the standard version. For the Llama-2-70B-chat-GPTQ, the GPTQ algorithm, as presented in \cite{frantar2022gptq}, was employed to quantize the Llama 2 models within AutoGPTQ library.

\subsection{LLMs performance evaluation}
Although additional information is available from the well documents, in this work, we focused only on the location (latitude and longitude) and depth (true vertical depth) of each well. We employed two metrics to assess the performance of our information extraction process. The first metric is the accuracy based on an exact match or $A_{EM}$, which is defined as:

\begin{equation}
    A_{EM} = \frac{N_{EM}}{N_{T}} \times 100\%
\end{equation}

where $A_{EM}$ denotes the accuracy; $N_{EM}$ and $N_T$ represent the number of entries that exactly match the true values and the total number of entries, respectively. In an ideal case, $A_{EM}$ would be 100\% if the LLM generates results that are exactly accurate, which may not always be the case in reality. Therefore, we also considered a second metric, the accuracy based on offset or $A_{OS}$, here. The rationale behind this metric is that it is also acceptable if a certain LLM generates a close approximation to the true value. For example, if the extracted location, in terms of latitude and longitude, is within 10 meters of the true location, we treat the result as correct. Similarly, if the extracted depth is within a range of 10 feet ($\sim$3 meter), we would accept this extraction result. From a practical point of view, an offset of 10 meters for the well location is considered acceptable because field operators can easily locate the well based on the extracted well location information. The definition of $A_{OS}$ is similar to that of $A_{EM}$ except $N_{EM}$ is changed to $N_{OS}$ as:

\begin{equation}
    A_{OS} = \frac{N_{OS}}{N_{T}} \times 100\%
\end{equation}

where $N_{OS}$ represents the number of entries that are within the offset threshold of the true value. In theory, for any information extraction task as presented in the work, $ A_{OS}$ should not be less than $A_{EM}$. In this paper, we calculate $A_{OS}$ and $A_{EM}$ for location and depth information extraction only rather than latitude, longitude, and depth. That is because the location can be represented as latitude and longitude. The location offset is calculated based on the geographical distance, also known as geodetic distance, which is the shortest arc length between two locations along the Earth's surface, using GeographicLib \cite{karney2015geographiclib} package. 

Our scope of this study is to propose a novel process for information extraction from well records by leveraging the capabilities of LLMs and testing this concept. Despite using only 160 well records for demonstration, the proposed information extraction workflow can rapidly process a significantly larger number of documents. The small dataset was used here because it enabled us to validate the approach quickly. Once a large dataset becomes available, it is anticipated that the developed framework can be easily scaled up for information extraction.

\section{Results}
\label{sec:results}
In this section, we demonstrate the capabilities of LLMs for information extraction using Llama 2 model. We begin by comparing the performance of Llama 2 70B model with various prompts. Next, we illustrate the performance differences among Llama 2 7B, 13B, and 70B models using the optimal prompt. Finally, we showcase the effectiveness of implementing the chain-of-thought strategy with the Llama 2 70B model and compare the performance with other LLMs, again using the selected optimal prompt.

\subsection{Prompt engineering for Llama 2 70B}
Before applying any LLM for information extraction, we need to formulate and then select the optimum prompts for the Question-Answering task. This can be achieved through prompt engineering, which involves designing proper prompts to achieve desired outcomes from LLMs \cite{liu2022design,reynolds2021prompt}. To find the best-performing prompt for our task of locating a well and identifying the depth of a well, we designed a total of four prompts ranging from Prompt 1 to 4, as shown in Table \ref{tab:table_vary_promt}. We started with a simple yet straightforward prompt and gradually increased the complexity of the prompts by providing more domain knowledge and detailed requirements. This additional information related to the specific documents of interest, would guide LLMs to extracting correct information of interest. Here, we evaluated the performance of Llama 2 70B using various prompts. 

Table \ref{tab:table_vary_promt} provides detailed information on four proposed prompts including prompt index, prompt content, and the corresponding explanation. Prompt 1 is the simplest one by just instructing LLMs to extract well information in terms of latitude, longitude, and depth information and to report in a JSON format. If users lack detailed information about the documents, the simple prompt can be used directly without extensive domain knowledge. On the other hand, Prompt 4 is the most comprehensive, ensuring that: (1) reported latitudes and longitudes are drilled latitudes and longitudes, and use decimal degrees as the unit; (2) longitudes are negative, given the well's location in the U.S.; (3) only true vertical depth is exported as depth information, despite that other depth information, e.g., measured depth, are available; (4) the true vertical depth is a positive number. By combining the proposed prompt with converted text through a text extraction process, a complete question was created for LLMs, which was then subjected to LLMs to perform information extraction. It is important to note that once the questions are formulated using the prompts, they can be directly utilized across various LLMs without any further adjustments.

\begin{table}[h]
\fontsize{10pt}{10pt}\selectfont
\caption{Different prompts for information extraction. The italicized texts in the second column highlights the differences between the current prompt and the previous one.}
\centering
\begin{tabularx}{\textwidth}{cp{0.5\textwidth}XX}
\toprule
        Prompt index & Prompt & Explanation \\ 
        \midrule
        Prompt 1 &  Extract the location using latitude and longitude, and well depth of the well described in this well completion report. Output only the latitude, longitude, and depth in JSON format as numbers, not strings, in a clean version. Only output the JSON and nothing else. Here is the OCR'd contents of the well completion report: & This prompt directs the LLM to extract the well's location (latitude and longitude) and depth from the well completion report, and to output the numbers in JSON format. This is the simplest prompt for this task. \\ \midrule
        Prompt 2 & \textit{Extract the drilled latitude (in degrees), longitude (in degrees), and true vertical depth (TVD) of the well described in this well completion report.} Output only the latitude, longitude, and depth in JSON format as numbers, not strings, in a clean version. Only output the JSON and nothing else. Here is the OCR'd contents of the well completion report: & 
        Provide more detailed instructions for reporting the well's location using decimal degrees, and outputting the well depth specifically in terms of True Vertical Depth (TVD). \\ \midrule
        Prompt 3 & Extract the drilled latitude (in degrees), longitude (in degrees), and true vertical depth (TVD) of the well described in this well completion report. \textit{Do not report depth in terms of Measured Depth (MD). Keep in mind that this text is extracted using optical character recognition (OCR), so the format may be jumbled. This well is in the western hemisphere, so the longitude should be negative.} Output only the latitude, longitude, and depth in JSON format as numbers, not strings, in a clean version. Only output the JSON and nothing else. Here is the OCR'd contents of the well completion report: & Additional instructions are provided to ensure the LLM not report measured depth, and the longitude should be negative given the location of the well of interest. \\ \midrule
        Prompt 4 & Extract the drilled latitude (in degrees), longitude (in degrees), and true vertical depth (TVD) \textit{(not footage at surface and not plug back total depth)} of the well described in this well completion report. Do not report depth in terms of Measured Depth (MD). Keep in mind that this text is extracted using optical character recognition (OCR), so the format may be jumbled. This well is in the western hemisphere, so the longitude should be negative. \textit{In addition, the true vertical depth cannot be negative.} Output only the latitude, longitude, and depth in JSON format as numbers, not strings, in a clean version. Only output the JSON and nothing else. Here is the OCR'd contents of the well completion report: & Ensure the LLM not using footage at surface and plug back total depth as well as depth information. Also, ensure that well depth information from the well completion report is not negative. If a negative value is found, it should be corrected to the corresponding positive value. This is the most complicated prompt for this task.\\
\bottomrule
\end{tabularx}
\label{tab:table_vary_promt}
\end{table}

\begin{figure}[h]
	\centering
        \includegraphics[width=6cm]{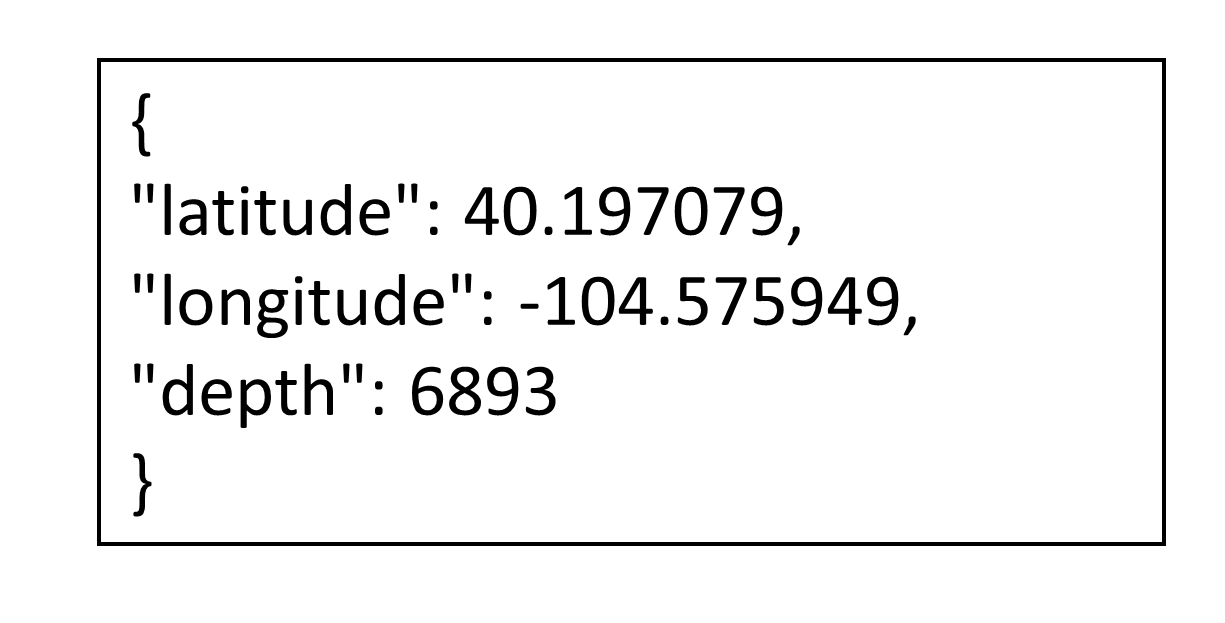}
	\caption{An example of information extraction output using Llama 2 70B.}
	\label{fig:llm_output}
\end{figure}

After running Llama 2 70B model with a question, an output as shown in Figure \ref{fig:llm_output} can be obtained. Figure \ref{fig:llm_output} represents the output from Llama 2 70B model for the drilling completion report in Colorado (in Figure \ref{fig:well_record_examples_for_this_paper_a}) using Prompt 1. As expected, the output is in the format of a JSON file within a Python environment. Specifically, the output for this example contains the names `latitude', `longitude', and `depth', along with their corresponding numbers as instructed by the prompt. In this example case, the latitude, longitude, and depth are 40.197079, -104.575949, and 6893 ft, respectively. The extracted information exactly matches the true values, demonstrating the good performance of Llama 2 70B model with Prompt 1. It should be noted that although JSON-style outcomes are generated by LLMs after analyzing the text-based well records, the current workflow does not have the capability to save the outputs directly as local JSON files. Therefore, an additional post-processing step is required to save the information extraction outputs as local files.

Let's now examine the final extraction performance for the four prompts across the 160 well records in Colorado and Pennsylvania. Given the fact that the well records from the U.S. are not in the same format, we here evaluate the performance of LLMs for information extraction separately. Again as introduced in Section~\ref{sec:method}, we used $A_{EM}$ and $A_{OS}$ as the metrics for performance evaluation. We reiterate that we used 10 meters and $\sim$3 meters (10 ft) as the thresholds for computing well location and depth offset, respectively. For the 150 Colorado drilling well record documents, the information extraction results obtained by Llama 2 70B with four different prompts are presented in Table \ref{tab:table_vary_promt_CO}.

Clearly, excellent location extraction performances were obtained using Llama 2 70B model for Colorado cases. As shown in this table, the values of $A_{EM}$ and $A_{OS}$ for location reach 100\% despite the varying contents of the prompt. This observation reveals that the locations of all 150 wells were correctly extracted from the well drilling completion reports in Colorado using the LLM and proposed prompts. In terms of well depth extraction, we find that Llama 2 70B model with Prompt 1 (the simplest prompt) yielded the lowest accuracy. Specifically, both $A_{EM}$ and $A_{OS}$ were 48\%, indicating that the locations of 72 documents were correctly identified. In other words, the Llama 2 70B model, when using Prompt 1, encountered difficulty in extracting information from the remaining 78 documents. Despite an offset of 10 ft was utilized for computing $A_{OS}$, the identical value in $A_{EM}$ and $A_{OS}$ illustrates the identified depth using the LLM with Prompt 1 deviates by more than 10 ft from the true value. For example, in one drilling well completion report, the actual well depth is 6806 ft, however, the extracted value is 17529 ft, which is about 10723 ft away from the true value. This observation illustrates that for this case, extracting well depth information was much more challenging than extracting well location information using Prompt 1. Except Prompt 1, Llama 2 70B with the remaining prompts resulted in great depth extraction performance with 100\% accuracy. This result illustrates that more comprehensive prompts (Prompts 2 to 4) facilitate information extraction more effectively than a basic prompt (Prompt 1).

Table \ref{tab:table_vary_promt_PA} compares well information extraction results for the 10 well records using Llama 2 70B with these predefined four prompts in Table~\ref{tab:table_vary_promt}. The major difference between the Pennsylvania and the Colorado case studies is that Llama 2 70B did not provide superior results for Pennsylvania. Clearly, none of the prompts resulted in completely correct extraction for the 10 Pennsylvania well records. For the location, Llama 2 70B model with the best-performing prompts (i.e., Prompt 4) resulted in a 70\% accuracy. This corresponded to 7 correctly extracted well locations. For the depth information extraction task, a value of 90\% was achieved for $A_{EM}$ and $A_{OS}$ from all the four prompts, indicating that we obtained correct depth information for 9 out of 10 documents. The relatively inferior extraction performance from Llama 2 70B for Pennsylvania is probably due to the low quality of text extracted with OCR for Pennsylvania well records. It is also interesting to see that, for the depth information extraction task, the performance in terms of $A_{EM}$ and $A_{OS}$ did not change with varying prompt contents. 

On the other hand, we observed that the extraction accuracy for locations increased with the complexity of prompts. For $A_{EM}$ and $A_{OS}$, Llama 2 led to 70\% accuracy using Prompt 4, compared to only 60\% accuracy for the remaining three prompts. The results in this case reveal that more complicated prompts result in better extraction performance. This general trend is confirmed by both $A_{EM}$ and $A_{OS}$ for extractions of location. Based on the information extraction results in Tables \ref{tab:table_vary_promt_CO}--\ref{tab:table_vary_promt_PA}, we see that a more detailed prompt often leads to better information extraction results. In the following sections of this paper, Prompt 4 was used for the investigation analysis. 

\begin{table}[h]
\caption{Information extraction results using Llama 2 70B model with different prompts for Colorado well completion reports.}
\centering
\begin{tabular*}{\linewidth}{@{\extracolsep{\fill}} ccccc}
\toprule
\multirow{2}{*}{Prompt index} & \multicolumn{2}{c}{Location}   & \multicolumn{2}{c}{Depth} \\
\cmidrule{2-5} 
            &  $A_{EM}$ & $A_{OS}$  &  $A_{EM}$ & $A_{OS}$ \\
    \midrule
    Prompt 1 & \textbf{100\%} & \textbf{100\%} & 48\% & 48\% \\
    Prompt 2 & \textbf{100\%} & \textbf{100\%} & \textbf{100\%} & \textbf{100\%}  \\
    Prompt 3 & \textbf{100\%} & \textbf{100\%} & \textbf{100\%} & \textbf{100\%} \\
    Prompt 4 & \textbf{100\%} & \textbf{100\%} & \textbf{100\%} & \textbf{100\%} \\
    \bottomrule
\end{tabular*}
\label{tab:table_vary_promt_CO}
\end{table}

\begin{table}[h]
\caption{Information extraction results using Llama 2 70B model with different prompts for Pennsylvania well records.}
\centering
\begin{tabular*}{\linewidth}{@{\extracolsep{\fill}} ccccc}
\toprule
\multirow{2}{*}{Prompt index} & \multicolumn{2}{c}{Location}   & \multicolumn{2}{c}{Depth} \\
\cmidrule{2-5}  
            &  $A_{EM}$ & $A_{OS}$  &  $A_{EM}$ & $A_{OS}$ \\
    \midrule
    Prompt 1 & 60\% & 60\% & \textbf{90\%} & \textbf{90\%}  \\
    Prompt 2 & 60\% & 60\% & \textbf{90\%} &  \textbf{90\%}  \\
    Prompt 3 & 60\% & 60\% & \textbf{90\%} &  \textbf{90\%} \\
    Prompt 4 & \textbf{70\%} & \textbf{70\%} & \textbf{90\%} &  \textbf{90\%} \\
    \bottomrule
\end{tabular*}
\label{tab:table_vary_promt_PA}
\end{table}

\subsection{Comparison of different Llama 2 models}
In this test, we used the best prompt via a zero-shot learning from the previous section to test the extraction performance of the three Llama 2 models, i.e., 7B, 13B, and 70B. Here, Prompt 4 was employed within Llama 2 7B and 13B to extract well location and depth information from the 160 well records. Subsequently, we compared these extraction results with those from Llama 2 70B model. Tables~\ref{tab:table_vary_model_CO}-\ref{tab:table_vary_model_PA} show the comparison results for Colorado and Pennsylvania cases, respectively. These two comparisons reveal that, in general, as the size (model parameters) of Llama 2 increases, better performance is achieved, though some deviations from the trend are observed, as shown in Table~\ref{tab:table_vary_model_CO}. For the 150 drilling completion reports in Colorado, Llama 2 7B model achieved an accuracy of 82.67\% accuracy in terms of $A_{EM}$ for depth, which is lower than the 90\% accuracy achieved by Llama 2 13B model. As expected, neither of the smaller models can surpass the 70B model in depth extraction. However, surprisingly, we find that for the location extraction task, the 7B model yielded a slightly better result when compared with the 13B for both $A_{EM}$ and $A_{OS}$, contrary to expectations. For example, the Llama 13B only achieved 64\% and 66\% for $A_{EM}$ and $A_{OS}$. Interestingly, the Llama 7B slightly outperformed it, achieving higher accuracy rates of 76.67\% for $A_{EM}$ and 77.33\% for $A_{OS}$. For the 10 Pennsylvania well records, we observed a consistent pattern: the larger model yielded better information extraction results. This conclusion is supported by data on both location and depth. For example, for the location information extraction, the Llama 2 7B, 13B, and 70B models resulted in accuracy values of 30\%, 60\%, and 70\% if $A_{EM}$ was used for evaluation, respectively. For the depth case, Llama 2 models with 70B and 13B parameters significantly outperformed the 7B model. The results presented here demonstrate that larger LLMs are generally more effective for information extraction tasks. It is recommended that users opt for larger models if they have the sufficiently powerful hardware support, as more advanced hardware is required to run larger LLMs.

\begin{table}[h]
\caption{Information extraction results using three Llama 2 models with Prompt 4 for Colorado well completion report.}
\centering
\begin{tabular*}{\linewidth}{@{\extracolsep{\fill}} ccccc}
\toprule
\multirow{2}{*}{Model name} & \multicolumn{2}{c}{Location}  & \multicolumn{2}{c}{Depth} \\
\cmidrule{2-5} 
    &  $A_{EM}$ & $A_{OS}$  &  $A_{EM}$ & $A_{OS}$ \\
    \midrule
    7B & 76.67\% & 77.33\% & 82.67\% & 82.67\% \\
    13B & 64\% & 66\% & 90.67\% & 97.33\%   \\
    70B & \textbf{100\%} & \textbf{100\%} & \textbf{100\%} & \textbf{100\%} \\
    \bottomrule
\end{tabular*}
\label{tab:table_vary_model_CO}
\end{table}

\begin{table}[h]
\caption{Information extraction results using three Llama 2 models with Prompt 4 for Pennsylvania well records.}
\centering
\begin{tabular*}{\linewidth}{@{\extracolsep{\fill}} ccccc}
\toprule
\multirow{2}{*}{Model name} & \multicolumn{2}{c}{Location}   & \multicolumn{2}{c}{Depth} \\
\cmidrule{2-5} 
    &  $A_{EM}$ & $A_{OS}$  &  $A_{EM}$ & $A_{OS}$ \\
    \midrule
    7B &  30\% & 30\% & 40\% & 40\% \\
    13B & 60\% & \textbf{70\%} & \textbf{90\%} &  \textbf{90\%}   \\
    70B & \textbf{70\%} & \textbf{70\%} & \textbf{90\%} &  \textbf{90\%} \\
    \bottomrule
\end{tabular*}
\label{tab:table_vary_model_PA}
\end{table}

\subsection{Impact of the chain-of-thought on the performance of Llama 2 70B}
\begin{figure}[h]
	\centering
        \includegraphics[width=\linewidth]{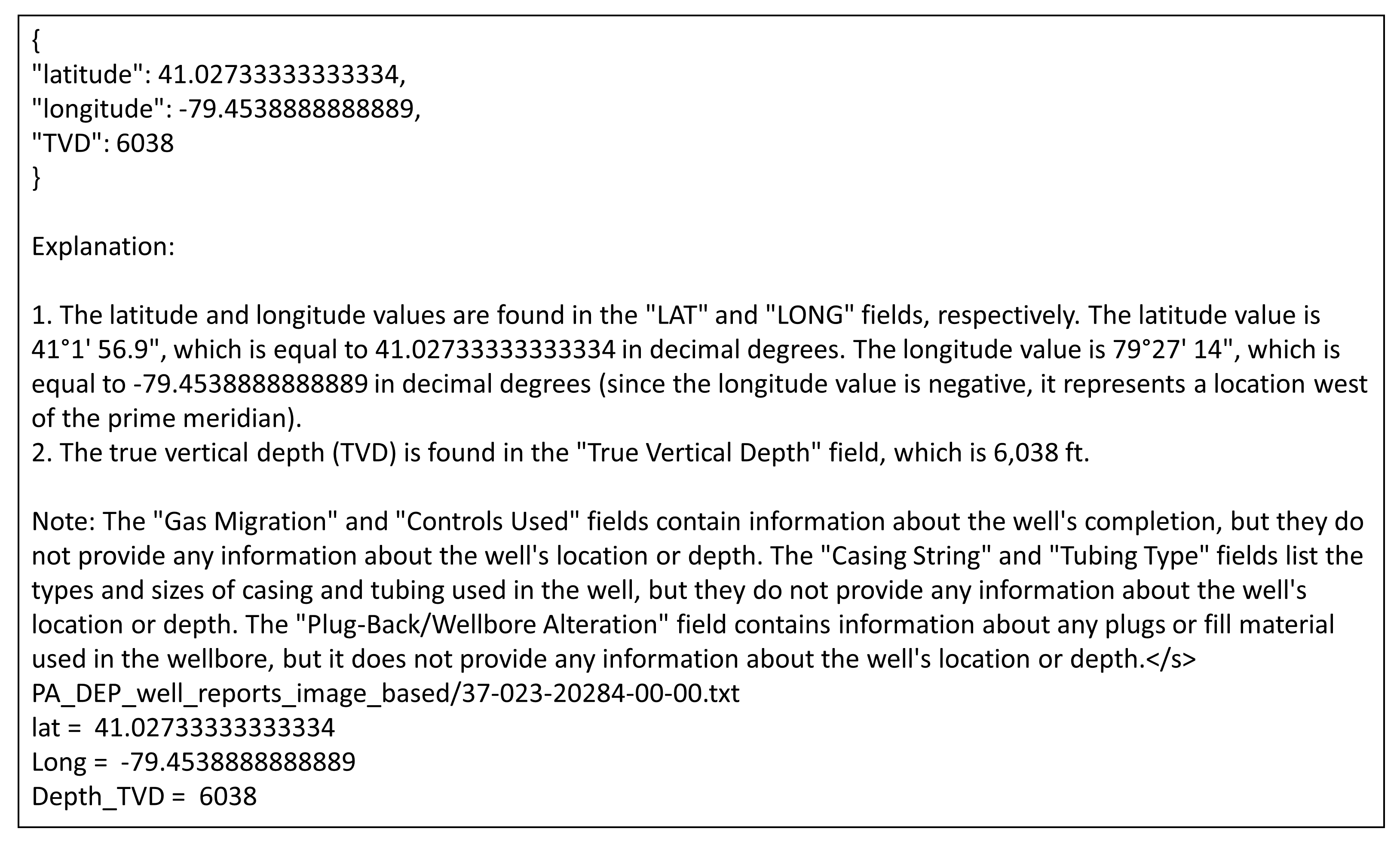}
	\caption{An example of LLM output for one Pennsylvania well record with implementing Chain-of-Thought. The true latitude, longitude, and depth are 41\textdegree1'56.9'', -79\textdegree27'14'', and 6038 ft, respectively. 
 }
	\label{fig:llm_output_pa_4_chain-of-thought}
\end{figure}

As presented in the previous sections, Llama 2, regardless of the prompt used or the size employed, was difficult to extract completely correct information from the 10 Pennsylvania well records. This is due to the unstructured nature of the text extracted from those well records with OCR. In this work, we also explored the possibility of enhancing extraction performance by incorporating the chain-of-thought approach with Llama 2 70B, using Prompt 4. Given the fact that the text extracted from the Colorado well drilling completion reports is more uniform, we focused only on those 10 Pennsylvania reports. We implemented the chain-of-thought approach to Prompt 4 in Table \ref{tab:table_vary_promt} by adding the following words to the end of the prompt: \textit{Please explain your detailed steps to get the numbers}. By incorporating this strategy, Llama 2 70B model generates the following output, as presented in Figure~\ref{fig:llm_output_pa_4_chain-of-thought}.

Unlike those without a chain-of-thought strategy, the texts generated by Llama 2 here exhibited more detailed `thinking' steps for extracting the numbers for location and depth. As shown in the texts, Llama 2 first identified the location in terms of latitude and longitude of 41\textdegree1'56.9'', -79\textdegree27'14'', respectively. The location numbers in degrees, minutes, and seconds accurately matched the true values in the original well record. However, as instructed by the prompt, the extracted location should use the unit of degrees, instead of degrees, minutes, and seconds. Therefore, another implicit task for the LLM is to convert the numbers from degrees, minutes, and seconds to decimal degrees, which needs a certain degree of mathematical skills. As shown in its output, Llama 2 model directly converted the latitude of 41\textdegree1'56.9'' to 41.02733333333334. However, the correct conversion should result in 41.032472. Despite the two values being very similar, a slight difference remained. Therefore, this example was not counted for $N_{EM}$ for calculating $A_{EM}$ because of the difference. The corresponding information extraction results with incorporating chain-of-thought strategy are presented in Table~\ref{tab:table_chain_of_thought_PA}. It is interesting to observe that, with the chain-of-thought employed, Llama 2 70B model resulted in the same or even worse extraction performance than that achieved without using the chain-of-thought strategy. For location extraction, Llama 2 70B with the chain-of-thought strategy yielded 60\% in terms of $A_{EM}$ and $A_{OS}$, which is lower than the 70\% by Llama 2 70B with no chain-of-thought strategy used. The comparison presented here reveals that the chain-of-thought strategy offered limited improvement for location and depth information extractions for Llama 2 70B. It is anticipated that if a post-processing procedure is applied to convert the location units from degrees, minutes, and seconds to decimal degrees, the results could be potentially improved. However, this is beyond the current scope of this paper. In addition, combining LLMs with external tools through the strategy of function calling may be one potential solution to this precise mathematical problem.

\begin{table}[h]
\caption{Comparison of information extraction results for Pennsylvania well records using Llama 2 70B Model with and without implementing of chain-of-thought strategy.}
\centering
\begin{tabular*}{\linewidth}{@{\extracolsep{\fill}} ccccc}
\toprule
\multirow{2}{*}{Scenario} & \multicolumn{2}{c}{Location}   & \multicolumn{2}{c}{Depth} \\
\cmidrule{2-5} 
    &  $A_{EM}$ & $A_{OS}$  &  $A_{EM}$ & $A_{OS}$ \\
    \midrule
    With Chain-of-thought &  60\% & 60\% & 90\% & 90\% \\
    Without Chain-of-thought & \textbf{70\%} & \textbf{70\%} & \textbf{90\%} &  \textbf{90\%} \\
    \bottomrule
\end{tabular*}
\label{tab:table_chain_of_thought_PA}
\end{table}

\subsection{Comparison with other LLMs}
In this work, we also tested the information extraction workflow using two additional models, including Mixtral 8$\times$7B and a commercial LLM model, namely GPT-3.5, from OpenAI (https://openai.com/). For the Mixtral 8$\times$7B, the result is presented in Table \ref{tab:table_Mixtral_CO_PA}. Interestingly, the Mixtral 8$\times$7B did not yield better information extraction result compared to Llama 2 70B model.

\begin{table}[h]
\caption{Information extraction results using Mixtral 8$\times$7B model with Prompt 4 for Colorado and Pennsylvania well reports.}
\centering
\begin{tabular*}{\linewidth}{@{\extracolsep{\fill}} ccccc}
\toprule
\multirow{2}{*}{State name} & \multicolumn{2}{c}{Location}   & \multicolumn{2}{c}{Depth} \\
\cmidrule{2-5} 
    &  $A_{EM}$ & $A_{OS}$  &  $A_{EM}$ & $A_{OS}$ \\
    \midrule
    Colorado     & \textbf{96.67\%} & \textbf{99.33\%} & \textbf{96.67\%} & \textbf{97.33\%} \\
    Pennsylvania & 30\% & 30\% & 50\% &  50\%   \\
    \bottomrule
\end{tabular*}
\label{tab:table_Mixtral_CO_PA}
\end{table}

For the GPT-3.5 model, we achieved better results than with all other models tested in this paper. Specifically, we obtained 100\% accuracy for all 150 Colorado documents and the 10 Pennsylvania documents (which are challenging to extract using other models). It was found that GPT-3.5 possesses superior mathematical capabilities to convert the unit's latitude and longitude. These findings underscore the advanced capabilities of GPT-3.5 in extracting information from historical documents, and illustrate the potential of more powerful models to provide higher accuracy. However, the commercial nature of GPT-3.5 limits its usage to an application programming interface (API) that incurs a cost for each use. 

\section{Discussions of challenges and opportunities}
\label{sec:discussions}
In this section, we briefly discuss the current challenges and potential opportunities of applying LLMs to tasks such as information extraction. This discussion is solely based on our results from extracting well location and depth data using 160 documents. In addition, given that the development of LLMs is progressing rapidly, it is possible that some of the information summarized here may not accurately reflect the latest advancements in LLMs.

\subsection{Enhance text conversion quality from historical documents}
As introduced previously, the current information extraction tasks require that the original historical documents be converted to texts before feeding into LLMs. This is because the LLMs employed in this work are designed to process textual inputs. Thus, the tasks heavily depend on the accuracy of the text extraction process used (e.g., OCR). However, even the best text extraction techniques still struggle to achieve 100\% accurate text conversions from documents such as PDFs and images. To deal with this challenge, it is recommended to further advance text extraction techniques to improve the accuracy and quality. Integrating computer vision techniques or machine learning algorithms could be potential areas. An alternative path to improving text extraction quality is to utilize multi-modal models that can extract textual information from the images directly. This is a promising avenue for future research.

\subsection{Improve the capabilities of LLMs}
The technology of LLMs advances rapidly in terms of new models, increased parameter sizes, and capabilities \cite{birhane2023science}. However, given that numerous LLMs are available from both the private and public domains, exploring other LLMs is necessary to get a better result. In this paper, we tested Llama 2 models. As discussed in Section \ref{sec:results}, Llama 2 model, despite the use of various prompts, changes in model parameter sizes, and the incorporation of the chain-of-thought strategy, could not achieve precisely correct information extraction from historical well documents. Therefore, it is worth testing other LLMs for the same task. 

Many other commercial online tools have been available for processing documents and information extraction, among which, Amazon Textract and Google Document AI are the two examples. We leveraged Google Document AI's Enterprise OCR for text extraction task in this study. However, Google Document AI offers additional capabilities for extracting information from unstructured or structured documents. Currently, the authors are exploring alternative approaches for extracting information from historical records with Google Document AI. In addition, given that these are cloud-based tools, users may need to consider data security issues when processing sensitive information data.

Another opportunity lies in fine-tuning the pre-trained LLMs for specific tasks. In this work, we focused solely on the zero-shot learning strategy, without performing any fine-tuning. However, fine-tuning LLMs could potentially be a better option, if feasible. Currently, we are investigating the improvement of fine-tuned LLMs for information extraction and will report their findings on performance in a future publication. Furthermore, it is also possible to incorporate multi-modal models for information extraction. Specifically, these models can directly take images or PDFs as inputs, eliminating the need for text conversion using OCR techniques. Although not employed in this study, it is also advisable to implement some post-processing procedures to enhance the information extraction performance.

\subsection{Overcome the challenges of extreme hardware requirements}
In order to use these LLMs offline, we must meet the hardware requirements, especially regarding GPUs due to the extremely large size of the LLMs. For example, according to the Hugging Face data repository, the total size of the standard version of Llama 2 70B-chat-hf is approximately 280 GB. Additionally, Hugging Face suggests using 4 $\times$ NVIDIA A100 GPUs for the deployment of Llama 2 70B models. While using the pre-trained LLMs as presented in this paper does not demand extensive computational resources, it still requires higher-end GPUs to run. In the information extraction here, we utilized an NVIDIA RTX A6000 GPU. Even with this GPU, we still have difficulty loading the full standard version of Llama 2 70B model. That was the reason why we used the quantized version of Llama 2 models for this study. Therefore, the extreme hardware requirements hinder the wide applications of LLMs. One way to address this challenge is through more advanced GPUs. Given the recent rapid advancements in GPU technology, the situation should continue to improve. Additionally, the development of smaller LLMs could also be a viable solution. Another alternative is to use commercial LLMs that are only available through an API.

\section{Concluding remarks}
\label{sec:concluding-remarks}
In this work, we presented an LLM-based workflow to extract vital information from well records for orphaned well management, including the well's location and depth. Data extraction from historic records is currently a labor-intensive process that takes time and is costly to perform. The information contained in well records is critically important for the successful plugging operations to reduce environmental impacts such as methane leakage from wellbores. To demonstrate the capability of information analysis workflow, we primarily focused on the publicly available Llama 2 model. To facilitate information extraction, we created multiple prompts, varying the instructions from the simplest to the most complex ones. Different variants of Llama 2 were also evaluated, including the 7B, 13B, and 70B models. Additionally, we also employed the chain-of-thought approach in an attempt to enhance performance. We tested the developed workflow using a dataset of 160 well records. Although this number is quite small, the goal of this paper is to prove the concept of this method. We emphasize that these forms are used only for validation of the approach, not for training the models. The development of an information extraction API capable of handling much larger datasets of well documented information is an ongoing project.

Several major conclusions are derived from the results. First, the content of the prompt impacts the final extraction results, even when an identical LLM is used. In this work, we found that Llama 2 70B model with Prompt 4 led to the best performance. The general trend is that the information extraction performance improves with the complexity of the prompt instructions. Therefore, it is recommended to optimize prompt content before using LLMs. Second, the size of the model is an important parameter that influences the results. With Llama 2 models, better performance was often obtained when a larger model was used. For example, Llama 2 70B model outperformed the smaller models, including the 7B and 13B variants. Third, although Llama 2 models achieved 100\% accuracy for the Colorado reports, they still had difficulties in correctly extracting information from some Pennsylvania well reports. This is because the quality of text extraction using OCR was low. Although Llama 2 70B extracted the correct location information in the units of degrees, minutes, and seconds after incorporating a chain-of-thought strategy, it did not accurately convert it into decimal degrees as instructed. 

While the developed workflow achieved good performance, especially for the PDF-based documents, opportunities for further improvement still remain. These include: (1) improving the quality of text extraction from historical documents, since the current workflow relies heavily on that; (2) fine-tuning the pre-trained LLMs for this specific task using a smaller dataset; (3) executing these information extraction tasks on higher-end hardware to enhance the results; (4) utilizing multi-modal models that can directly process PDFs and images, thereby eliminating the need for text extraction; and (5) utilization of LLM function calling techniques to aid the LLM with routine tasks like converting. These techniques have the potential to automate much of the information extraction workflow, accelerating the plugging of abandoned wells and enabling large scale data collection for research purposes.

\section{Acknowledgements}
This work is supported by the Department of Energy's Undocumented Orphan Well program through the CATALOG consortium (catalog.energy.gov). LA-UR number 'LA-UR-24-23837'.

\bibliographystyle{ieeetr}
\bibliography{references}  
\end{document}